\documentclass[12pt]{article}
\usepackage{graphics}
\newcommand{\la}[1]{\label{#1}}

\newcommand{\be}{\begin{equation}}
\newcommand{\ee}{\end{equation}}
\newcommand{\ba}{\begin{eqnarray}}
\newcommand{\ea}{\end{eqnarray}}
\newcommand{\bastar}{\begin{eqnarray*}}
\newcommand{\eastar}{\end{eqnarray*}}

\title{Are Glueballs Knotted Closed Strings?  \\ }\vskip 0.4cm 
\author{ \rm Antti \ J. \ Niemi\thanks{Antti.Niemi@teorfys.uu.se} 
\hspace{0.1cm}
\\ \\ 
{\it Department of Theoretical Physics, Uppsala University,} \\ 
{\it Box 803, S-75 108 Uppsala, Sweden} }

\begin{document}

\maketitle

\begin{abstract}
\noindent
Glueballs have a natural interpretation as closed strings
in Yang-Mills theory. Their stability requires that the 
string carries a nontrivial twist, or then it is knotted.
Since a twist can be either left-handed or right-handed,
this implies that the glueball spectrum must be degenerate.
This degeneracy becomes consistent with 
experimental observations, when we identify  
the $\eta_L(1410)$ component of the $\eta(1440)$ pseudoscalar
as a $0^{-+}$ glueball, degenerate in mass with the widely 
accepted $0^{++}$ glueball $f_0(1500)$. In addition of 
qualitative similarities, we find that these two states also share 
quantitative similarity in terms of equal production ratios,
which we view as further evidence that their structures 
must be very similar. We explain how our string picture 
of glueballs can be obtained from Yang-Mills
theory, by employing a decomposed gauge field. We also
consider various experimental consequences of our proposal,
including the interactions between glueballs and quarks
and the possibility to employ glueballs as probes
for extra dimensions: The coupling of strong interactions 
to higher dimensions seems to imply that absolute color
confinement becomes lost.
\end{abstract}
\vfill\eject
\noindent
Lord Kelvin was first to suggest that knotted strings 
could have fundamental importance. In \cite{kel1} he proposed that
atoms, which at the time were considered as elementary particles, could be
interpreted as knotted vortex tubes in aether. Subsequently he also
conjectured \cite{kel2} that vortex filaments in the shape of torus knots
should be stable. Kelvin's theory of vortex atoms has long ago subsided.
However, at the time it was taken seriously which led to an extensive study 
and classification of knots. In particular the results obtained by Tait
\cite{tait} remain a classic contribution to mathematical knot theory.

Presently, it is commonly accepted that fundamental interactions are
described by string theories \cite{green}, with different elementary
particles corresponding to the vibrational excitations of a primary
string. But the relevance of nontrivial string topology such as
an appropriate extension of the concept of a knot in a higher 
dimensional string (membrane) has until now been studied sparsely.
However, there are some obvious connections between string theory 
and knot theory, for example the Chern-Simons action is related both 
to conformal field theories and three dimensional knot 
invariants \cite{wit1}. Hopefully one day an intimate relationship 
between knots and nonperturbative strings will be established. 

Even though the relevance of knots at the level of
a fundamental string theory remains open, in strong interaction
physics we have good reasons to believe that knotted strings are present.
Indeed, it is widely accepted, and also confirmed by detailed
lattice simulations, that in QCD the collective excitations of gluons lead 
to the formation of a confining string between two 
widely separated quarks \cite{kuti}. Consequently it is only 
natural to expect that {\it glueballs} could be interpreted 
as closed configurations of this string, emitted 
{\it e.g.} by a fluctuation in a (relatively) long open 
string that connects two widely separated quarks.
Since glueballs relate directly to such fundamental issues
in strong interaction physics as confinement and the formation
of mass gap in Yang-Mills theory, their theoretical and
experimental study is centrally important for our 
understanding of the properties of hadronic particles, 
the source of most known mass in the Universe. 

A homogeneous and structureless linear string has an energy which 
is directly proportional to its length. In the absense of other contributions 
to the energy besides the linear tension, a closed string that has for
example a toroidal shape (unknot) is then unstable, it shrinks 
away by minimizing the length. This is what happens to closed vortex 
loops for example in the abelian Higgs model with a single complex 
field, an isolated closed toroidal vortex ring in type-II superconductor
is unstable against shrinkage.
 
But if a glueball corresponds to a closed string of gluonic flux,  
it must be stable within the Yang-Mills part of QCD. This is a direct
consequence of a mass gap, which prevents free massless gluons from appearing
in the spectrum. Consequently in the absence of quarks, or in the limit
of heavy quark masses, there is nothing 
to which a closed gluonic fluxtube could decay into.
This means that in the case of Yang-Mills theories there must 
be additional contributions to the energy of a closed gluonic string 
besides the linear string tension. Since lattice simulations indicate that a 
straight, linear gluonic string can only be subject to a (classically) linear 
tension \cite{kuti}, any additional force which stabilizes 
{\it e.g.} a closed toroidal string against shrinkage must have a geometric 
origin, it is present in a toroidal configuration but absent from 
a straight linear string. The natural source for such a stabilizing force 
is in the three dimensional extrinsic geometry of the fluxtube, 
the way how it twists and bends. In fact, suppose that we
bend a finite length linear piece of a fluxtube so that it forms
a toroidal ring {\it i.e.} an unknot. If we twist the fluxtube 
once around its core before joining the ends, we can expect that there is a 
centrifugal twist contribution to the energy which ensures the stability 
of the toroidal fluxtube against shrinkage \cite{nature}. Indeed, from general 
considerations \cite{landau} which can be interpreted as amounting 
to universality arguments in the sense of renormalization groups,
we expect that the leading order contribution to the energy from the 
bending and twisting of a linear fluxtube scales inversely to the
length $L$ of the fluxtube. Together with
the linear tension we then expect 
in the limit of long closed fluxtubes \cite{stroga}
\[
E \ = \ \sigma L \ + \ \frac{a}{L}
\]
where $\sigma$ is the (linear) string tension, and $a \geq 0$ is a parameter
characterising the (average) twisting and bending contributions to the energy.
From this we find that the energy of the fluxtube attains its 
minimum at a
non-vanishing value of $L$, 
\[
L_{min} \ = \ \sqrt{\frac{a}{\sigma}}
\]
This means that a closed, possibly knotted string becomes stabilized
against shrinkage, with a finite value of length $L_{min}$. 

But the present consideration also reveals an obvious degeneracy 
in the energy spectrum of stable closed fluxtubes. Namely,
before we join the ends of an open fluxtube to form the closed
configuration, the twist that we introduce can be either 
left-handed ($L$) or right-handed ($R$). In the absence 
of any parity symmetry breaking terms, we then have two different but stable 
configurations corresponding to the left-twisted and the right-twisted 
toroidal string (unknot configuration). If Yang-Mills theory is 
indeed capable of describing left-handed and right-handed twist, 
in the absence of any parity violating term in the Hamiltonian we 
expect that in the quantum theory we have a twofold mass-denegeracy
in the eigenstates of the Yang-Mills Hamiltonian, 
the left-handed twisted closed fluxtube $|L>$ and the right-handed 
twisted closed fluxtube $|R>$. Specifically, a
parity operator $P$ relates these two quantum states, 
\[
P|L> = |R> \ \ \ \ \ \ \ {\rm and} \ \ \ \ \ \ \  
P|R> = |L>
\]
and we can diagonalize $P$ by setting
\[
|\pm> \ = \ \frac{1}{\sqrt{2}}( |L> \pm |R> )
\]
with
\[
P |\pm> \ = \ \pm | \pm >
\]
and if
\[
[P, H_{ym} ] \ = \ 0
\]
the energy spectrum of the Yang-Mills hamiltonian $H_{ym}$ 
is then $P-$degenerate. In particular,
for each parity-even closed string there must be a parity-odd
closed string. This means
that {\it if glueballs are closed, knotted fluxtubes which are stable
within the pure Yang-Mills part of QCD, they must reflect the
two-fold twist-parity degeneracy \cite{ulr}.}   

A parity degenerate glueball spectrum in a Yang-Mills theory
can actually 
be considered natural in the following sense: At very high energy the 
Yang-Mills theory describes asymptotically free 
massless gluons. These gluons can be polarized in two different ways,
they can be chosen to be either left-handed polarized or
right-handed polarized. But this implies a doubling in the
spectrum which is not akin the doubling we obtain 
by twisting the string in either left-handed or right-handed manner. 
Indeed, when we move towards the low energy phase,
if the polarization doubling survives the ensuing phase transition
it is only natural that it is reflected in a left-right 
doubling of the glueball spectrum.

Unfortunately, such a parity degeneracy in the glueball spectrum
is not consistent with lattice computations, which predict a 
highly non-degenerate spectrum \cite{mor}. Specifically, lattice
computations suggest that the lowest energy glueball is a
$J^{PC} = 0^{++}$ state with a mass of $1611 \pm 163 \ MeV$, 
while the parity transformed $J^{PC} = 0^{-+}$ state has a 
mass which is in the vicinity
of $2500 \ MeV$. This is even 
above the lattice prediction of the 
mass of $2^{++}$ glueball which is about
$2232 \pm 310 \ MeV$. 

In lattice simulations there does not appear to
be any direct evidence why glueballs should be interpreted
as stable closed strings. Consequently there is no intrinsic reason for 
a parity doubling in the spectrum. But the interpretation of a 
glueball as a closed string is not inconsistent with lattice 
computations, maybe one just does not know how to describe a twisted
closed gluonic string on a lattice. Thus it could simply   
be, that this aspect of glueballs has not yet emerged. 
But if lattice computations will eventually identify glueballs
as closed gluonic strings which is quite possible, we still
have the puzzle 
to explain the stability of a glueball in the infinite quark
mass limit in a manner which is consistent with the computed,
apparently highly asymmetric mass spectrum with respect to twisting.

In the light of these lattice computations, at the moment it appears 
to be very brave to conclude that our proposal to interpret
glueballs as twisted closed strings is theoretically viable. 
But when we make comparisons with experimental results, we 
arrive at a conclusion that there is nothing apparently 
inconsistent with our proposal \cite{ulr}. For this we note that
there is  a wide experimental agreement on the reliability of lattice 
computations to the extent, that the lowest mass glueball 
$0^{++}$ should have a mass which is about the same as the
mass of isoscalar $q \bar q$ mesons. Since these can be grouped into SU(3) 
flavor nonets any candidate for a glueball should appear as 
a supernumerous state. In addition, a glueball candidate
should be consistent with the following experimental 
signatures \cite{pdg}:   
 
\vskip 0.4cm

{\it i)} Glueballs should be produced in $p \bar p$ annihilation
processes, since a $q \bar q$ annihilation leads to a gluon-rich
environment which strongly favors the formation of gluonic 
degrees of freedom. In fact, due to the LEAR experiment the 
nucleon-antinucleon annihilation processes are now the 
dominant source of data for studying the glueball formation.

{\it ii)} The central region of various other high-energy hadron-hadron 
scattering processes should similarly favor the production of 
glueballs since it is neither populated by target nor by beam quarks.

{\it iii)} The radiative decay of quarkonium states and 
especially the radiative J/${\psi}$ decay should be a prime 
source of glueballs.

{\it iv)} The decay branching fractions of a glueball should be
incompatible with SU(3) predictions for $q \bar q$ states.
 
{\it v)} Since glueballs have no direct coupling to electromagnetism, 
they should be absent in any $\gamma\gamma$ collision processes.

\vskip 0.4cm

The relatively narrow state
$f_0(1500)$ is widely considered as
the best available candidate for the lowest 
mass $0^{++}$ glueball \cite{pdg}. It does fullfill all of the experimental 
criteria {\it i)-v)}, and it also has a mass which fits well
the lattice prediction for the $0^{++}$ glueball state.

The lattice predicts further, that the next glueball
should be a $2^{++}$ state with a mass of $2232\pm 
\ 310 MeV$ \cite{mor}. This prediction seems to fit in a wonderful manner
the observation of $f_J(2220)$, which has been for this reason
proposed as a candidate for the $2^{++}$ glueball \cite{pdg}. 
But with new data from $p \bar p$ annihilation processes
the existence of a $f_J(2220)$ state is in serious doubt, it
is fading away \cite{pdg}. Consequently at the present there 
does not seem to be any widely accepted candidate for 
the $2^{++}$ glueball, with a mass which is consistent 
with the computed lattice glueball spectrum.

While the lattice prediction of the $0^{++}$ glueball appears
to be almost perfect, the apparent difficulty of lattice
to predict the mass of a $2^{++}$ glueball could indicate that
a full understanding of glueball spectrum remains to be achieved.  
Indeed, at the moment our experimental observations
on the glueball spectrum are also quite lacking
and marred with puzzling observations. Notorious in this respect is 
the $\eta(1440)$, a $J^{PC} = 0^{-+}$ pseudoscalar 
which is most likely a mixture of two particles, 
the $\eta_H(1480)$ and the $\eta_L(1410)$ \cite{pdg}

The $\eta_H(1480)$ couples strongly to kaons. It seems to be
an almost ideal mix of $s \bar s$ which makes
it a likely candidate for the $s \bar s$ member of the pseudoscalar
nonet. But this implies that the $\eta_L(1410)$ becomes
supernumerous in this nonet. Consequently it is a 
candidate for exotics: In a full analogy with
the nearby $f_0(1500)$, the $\eta_L(1410)$ is a relatively narrow 
state that fullfills {\it all} of the above 
criteria {\it i)-v)} expected from a glueball. But since 
lattice computations predict that the mass of a $0^{-+}$ glueball 
should be in the vicinity of $2500 \ MeV$, in the literature 
\cite{glennys} this has been used as an argument to 
suggest that $\eta_L(1410)$ could be 
a new degree of freedom which is beyond the Standard Model QCD. 
According to \cite{glennys} it could even relate 
to a bound state of light gluinos. 

Here we consider a much more conventional possibility, with
no need to go beyond the Standard Model, by interpreting
the $\eta_L(1410)$ as the $J^{PC} = 0^{-+}$ glueball state \cite{ulr}. 
In fact, we propose that in the picture where we view glueballs as stable 
closed strings, the $\eta_L(1410)$ has a natural interpretation as
the parity-odd partner of the $0^{++}$ glueball corresponding
to $f_0(1500)$. Consequently the experimental 
investigation of the $\eta_L(1410)$ state becomes a direct test 
of our theoretical understanding of fluxtube structures in QCD. 

We note that the slight $90 \ MeV$ mass difference 
between the $\eta_L(1410)$ and the $f_0(1500)$ is
quite insignificant. Theoretically, this mass 
difference could signal the presence of a 
small CP-violating $\theta$-term in the QCD action. 
Experimentally, it is also well known that interferences 
with nearby resonances can easily shift masses by as much as 
$5\%$. Consequently the quoted values for the $f_0(1500)$, 
which are in the range of $1445-1560 \ MeV$ dependending on the 
reaction channels, are not necessarily inconsistent with a 
mass degeneracy with $\eta_L(1410)$. In fact, since $f_0(1500)$ is 
known to be heavily mixed with nearby meson states a precise
experimental mass comparison becomes quite delicate.

Besides similarity in mass and in the qualitative criteria {\it
i)-v) } which are also all satisfied by $\eta_L(1410)$, 
the $f_0(1500)$ and $\eta_L(1410)$ have also
additional quantitative similarities,
such as production ratios. These are clear indicatives that
the two states must have very similar internal structures. 
For example, in $p \bar p$ annihilations 
the production of $f_0(1500)$ has the
following measured branching ratios \cite{CB2}: 
\[
{\tt BF}\left( \frac{ p \bar p \rightarrow f_0(1500) }{ 
p \bar p \rightarrow 3\pi^0 }\right) \ = \ (13 \pm 4) \%
\]
\[
{\tt BF}( p \bar p \rightarrow 3\pi^0) \ = \ (5.5 \pm 1.0) \times 10^{-3}
\]
When we take into account that $2\pi^0$ decays represent $(9.3 \pm 2.5)\%$ 
of all $f_0(1500)$ decays \cite{AMS98} we conclude that 
\be
{\tt BF}(p \bar p \rightarrow f_0(1500)) \ =  \ (7.7 \pm 3.8) \times 
10^{-3}
\la{br1}
\ee
In our annihilation rate into $3\pi^0$ 
we have accounted for the latest value for the annihilation into 
$2\pi^0$ from the Crystal Barrel Collaboration \cite{CB3}. But we 
also note that the OBELIX collaboration quotes a value which is more than a 
factor of 2 smaller. There is also a disagreement that 
influences the normalization of our production ratio \cite{ratio1}, 
\cite{ratio2}. 
 
For $\eta_L(1410)$, the $K\bar K\pi$ and $\eta\pi\pi$ decay modes are 
expected to be the dominant \cite{glennys}. Both have been measured in
$p \bar p$ annihilations, with results \cite{CB4}
\[ 
{\tt BF} \left( \frac{ p \bar p \rightarrow \eta_L(1410)\pi\pi }
{ \eta_L(1410) \rightarrow \eta\pi\pi} \right) \ = \ (3.3 \pm 1.0) 
\times 10^{-3}
\]
and \cite{bai}
\[
{\tt BF} \left( \frac{ p \bar p \rightarrow \eta_L(1410)\pi\pi } 
{ \eta_L(1410) \rightarrow K\bar K\pi}\right) \ = \
(2.0 \pm 0.2) \times 10^{-3}
\]
When we add these, we find 
\be
{\tt BF} ( p \bar p \rightarrow \eta_L(1410)\pi\pi )
\ = \ (5.3 \pm 1.7) \times 10^{-3}
\la{br2}
\ee
In comparing with (\ref{br1}) we conclude that the $p \bar p$
annihilation production rates for our two glueball candidates
$f_0(1500)$ and $\eta_L(1410)$ in $p \bar p$ are remarkably equal,
when accounting for experimental uncertainties.

We have also compared the observed production rates in 
radiative J/$\psi$ decays. The branching fraction for the production 
of the $f_0(1500)$ and its subsequent decay into $4\pi$ has been 
measured to be $(8.2 \pm 1.7) \times 10^{-4}$ \cite{bug}, with $4\pi$
decays accounting for $61.7 \pm 9.6 \%$ of all $f_0(1500)$ decays. 
Therefore, we expect a branching fraction of
\[
{\tt BF }( J/\psi \rightarrow \gamma f_0(1500)) \ = \ (1.3 \pm 0.3) 
\times 10^{-3} 
\]
The branching fraction for the production of the $\eta_L(1410)$ 
can be determined again from its decays into $K\bar K\pi$ 
\cite {Mark1} and $\eta\pi\pi$ \cite {Mark2}. Adding the results 
leads to a branching fraction
\[
{\tt BF} (J/\psi \rightarrow \gamma \eta_L(1410))
\ = \ (1.0 \pm 0.46) \times 10^{-3}
\]
Again, we find that in the case of radiative 
J/$\psi$ decays the production rates of $f_0(1500)$ and
$\eta_L(1410)$ are remarkably similar 
within experimental uncertainties.

We find that these similarities in the rates, 
the relative narrowness of both $f_0(1500)$ and $\eta_L(1410)$,
and the fact that both satisfy all of the overall criteria $i)-v)$ 
are quite remarkable. To us this strongly suggests that these two
states must have very similar internal structures. Since there is
wide agreement to interpret $f_0(1500)$ as a $0^{++}$ glueball, 
the natural interpretation of $\eta_L(1410)$ is then in terms 
of a $0^{-+}$ glueball. As it is furthermore (essentially) 
degenerate in mass with $f_0(1500)$, 
the natural explanation is that these two states
are the parity even and odd linear combinations of closed 
toroidal strings with a single left-handed and right-handed twist
\cite{ulr}.  
Indeed, the present proposal to identify $\eta_L(1410)$ as a
$0^{-+}$ glueball which is denegerate in mass with the $0^{++}$ 
glueball $f_0(1500)$ is clearly 
very consistent with the picture that glueballs
correspond to closed strings. However since our proposal to
interpret $\eta_L(1410)$ within the Standard Model is somewhat 
unorthodox and our conclusion that it is a glueball is not consistent
with lattice predictions, there is an obvious need for some 
theoretical backing. For this we shall now explain how our model 
of glueballs could be supported by the structure of a Yang-Mills theory,
and in particular how the presence 
of the twisting-degree of freedom and the formation of
closed fluxtubes can be realized.

In order to identify the twisting degree of freedom 
directly in the Yang-Mills theory, we employ
a decomposition of the gauge field $A^a_\mu$ \cite{lud1}-\cite{lud3}.
The decomposition of vectors and tensors in terms of 
their irreducible components is a common problem
in Physics. For example, it is widely employed in fluid dynamics 
where the velocity three vector decomposes into its
gradient and vorticity components, and in classical
electrodynamics the four dimensional
Maxwellian field strength tensor $F_{\mu\nu}$ becomes 
dissected into its electric and magnetic components. 
For notational simplicity but without any loss of generality
we consider here a decomposition of $A^a_\mu$ in 
a SU(2) Yang-Mills theory. This decomposition \cite{lud3}
interprets the Cartan component
$A^3_\mu$ as a U(1)$\ \in \ $SU(2) gauge field. 
Then 
\[
A^+_\mu = A^1_\mu + i A^2_\mu
\]
together with its complex conjugate transforms
as charged vector field under the corresponding U(1)
gauge transformation, corresponding to diagonal SU(2) gauge 
rotations. The two vectors $A^1_\mu$ and $A^2_\mu$ are in
a plane in a four dimensional space. We parametrize
this plane using a {\it twobein} ${e^a}_\mu$ ($a=1,2$) with
\[
{e^a}_\mu {e^b}_\nu = \delta^{ab}
\]
We define 
\[
{\bf e}_\mu = 1/\sqrt{2}({e^1}_\mu + i {e^2}_\mu)
\]
With these, we can then represent
the most general $A^+_\mu$ as
\[
A^1_\mu + iA^2_\mu \ = \ i \psi_1 {\bf e}_\mu + i \psi_2
{\bf e}^\star_\mu
\]
Here $\psi_1$ and $\psi_2$ are two complex fields; see 
\cite{lud3} for details. 
We define 
\[
\rho^2 = |\psi_1|^2 + |\psi_2|^2
\] 
and introduce a three-component unit vector
$\vec n$ by
\be
\vec n \ = \ \frac{1}{\rho^2} ( \psi_1^\star \ \psi_2^\star)
\vec \sigma \left( \matrix{ \psi_1 \cr \psi_2 } \right)
\la{defn}
\ee
Here $\vec \sigma$ are the standard Pauli matrices.
We substitute the ensuing decomposed fields in the Yang-Mills
action, using the background gauge condition
\be
D^{ab}_\mu(A^3) A^b_\mu = 0
\la{gauge}
\ee
{\it w.r.t.} the Cartan component $A^3_\mu$.
When we keep {\it only} terms which involve $\rho$
and $\vec n$, we get \cite{lud3}, \cite{edw}; \cite{ludvig}
\be
- \frac{1}{4} \int d^4x F_{\mu\nu}^2
= \int d^4x \left\{ (\partial_\mu \rho)^2 + 
\rho^2 (\partial_\mu n^a)^2 + \frac{1}{4} ( \epsilon_{abc}
n^a \partial_\mu n^b \partial_\nu n^c)^2 + V(\rho^2, \vec h \cdot
\vec n) + \dots \right\}
\la{act}
\ee
Here $V$ is a potential term that involves some of the 
additional terms that we have deleted. This potential
term breaks the global $O(3)$ invariance of the first three
terms on the {\it r.h.s.} under rigid rotations of 
the vector $\vec n$. This is necessary, as it gives a
mass to the two Goldstone bosons corresponding to the
breaking of the global $O(3)$ by boundary condition
to $\vec n$ at spatial infinity \cite{lud3}. 

The function $\rho$ can be interpreted as the average density of the
two scalars $\psi_1$ and $\psi_2$,
\be
\int d^4x \ (A^1_\mu A^1_\mu + A^2_\mu A^2_\mu) 
\ = \ \int d^4x \ (|\psi_1|^2 + |\psi_2|^2) \ = \ \int d^4x 
\ \rho^2
\la{zakh}
\ee
This is {\it a priori} a gauge dependent quantity, but
when we minimize (\ref{zakh}) along gauge orbits the
minimization selects {\it exactly} the background gauge condition 
(\ref{gauge}). This means in particular, that the minimum value
$\rho_{min}$ of (\ref{zakh}) is a gauge invariant quantity
\cite{zakh}. A one-loop computation suggests that 
this minimum value $<\! \rho_{min} \! >$ is nonvanishing 
\cite{lisa}.

For $<\rho^2> \not= 0$ the action (\ref{act}) is known to 
support stable, knotted solitons \cite{nature}, \cite{hieta}. 
The simplest soliton
solution is either a left-handed or a right-handed unknot {\it
i.e.} a twisted, doubly-degenerate state as we need for
our picture of glueballs.

The knottedness and in particular the twisting of the soliton
relates to the topology of the unit vector field $\vec n$. For a finite
energy configuration, $\vec n$ must go to a constant vector
$\vec n \to \vec n_0$ at large distances, with $\vec n_0$
a minimum of the potential $V(\rho^2, \vec h \cdot
\vec n)$ at spatial infinity. Consequently $\vec n$ defines a mapping
from the compactified $R^3 \sim S^3 \to S^2$. These mappings fall into
nontrivial homotopy classes $\pi_3(S^2) \sim Z$ and can be 
characterized by the Hopf invariant \cite{nature}
For this we introduce the two-form
\[
F \ = \ < d \vec n \wedge d \vec n \ , \ \vec n >
\]
Since the cohomology group $H_2(S^3) = 0$, the preimage of $F$ on
$S^3$ is exact and we can write 
\[
F = dA
\]
The Hopf invariant $Q_H$ is then given by
\[
Q_H \ = \ \int F \wedge A
\]
It computes the self-linking number of a knotted stringlike
soliton, and in particular describes its self-twisting. 
Provided the knotted solitons of (\ref{act}) indeed survive 
in the quantum Yang-Mills theory, we then have natural 
candidates for the glueballs as knotted as twisted,
$L-R$ degenerate fluxtubes.

Obviously it would be too {\it naive} to expect that
the solitons of (\ref{act}) actually 
provide a {\it quantitatively} accurate description 
of glueballs. For this we must account for the roughening which is  
due to quantum fluctuations in the additional fields that 
appear in the pertinent decomposition of $A^a_\mu$ in the 
full SU(3) Yang-Mills theory. However, it is interesting 
to consider some of the predictions of these solitons.
Maybe some of their properties are sufficiently universal 
to survive a more comprehensive analysis in the full SU(3) theory. 

The energy spectrum of the solitons in (\ref{act}) follows a 
rational curve in their self-linking number (Hopf invariant)
$Q_H$ \cite{vaku}:
\[ 
E_Q \ \geq \ c \cdot |Q_H|^{3/4}
\]
If $f_0(1500)$ and $\eta_L(1410)$ are indeed the lowest mass
states, we then have $c \approx 1500 \ MeV$ which suggests the 
mass spectrum
\be
M_Q \ \approx \ 1500 \cdot |Q_H|^{3/4} \ MeV 
\la{mass}
\ee
which predicts that the next ($|Q_H|=2$) glueball has
a mass in the vicinity of $2500 - 2600 \ MeV$. This
value is well within the range of the planned radiative 
charmonium decay experiments at CLEO-II and BES. 
Even though we expect this estimate to be very crude,  
there are general model independent, topological
arguments \cite{bol2} which suggest that the 
$3/4$-scaling law should describe a universal behaviour, 
in the limit of large values $|Q_H|$. If this universality 
extends to smaller values of $|Q_H|$, there could be very interesting 
physics around $6.4 - 6.5 \ GeV$: According to (\ref{mass}) this energy
corresponds to $Q_H = \pm 7$, and numerical simulations \cite{hieta}
suggest that the ensuing soliton is a trefoil which is a 
{\it chiral} knot. Consequently, besides the twisting we have an
additional discrete symmetry with ensuing mass degeneracy, corresponding
to the two-fold mirror symmetry of the trefoil.
Curiously, we note that this mass value is near the 
upper reach of energy at the recently approved antiproton 
facility at GSI.

The $3/4$ scaling-law of energy also suggests that the interaction
between two knots is attractive. In particular, two singly-twisted
unknots can form a bound state which is a doubly-twisted unknot. 
This could eventually develop into a probe on the properties of 
glueballs, if indeed described by knotted fluxtubes.

In three spatial dimensions the self-linking
number of a knot is a topological invariant.
But if a knot becomes embedded in a space with more than
three spatial dimensions its self-linking number ceases to be
a topological invariant, and the knot can
disentangle. This suggests that the stability and decay
properties of glueballs could be developed to an experimental
test, to explore the dimensionality of space-time as
seen by strong interactions. In particular, in three space
dimensions the lowest energy glueball can only decay by the
formation of quark-antiquark pairs {\it i.e. } into mesons.
But the presence of extra dimensions suggests that there
are additional, purely gluonic decay channels where the glueball
could decay by unknotting itself thru higher dimensions, thus
not necessarily decaying into quarks. On the other hand,
the presence of a mass gap in four dimensional Yang-Mills theory
excludes the presence of massless gluons. Consequently 
the existence of extra dimensions could mean that
absolute confinement of gluons is lost. 

Indeed, in full four dimensional QCD we expect that 
(virtual) $q \bar q$ pairs instabilize a closed and 
knotted gluonic string: A closed string opens itself and 
becomes disentangled into another closed string but with a 
different self-linking number, through the 
formation and subsequent annihilation of a quark-antiquark pair. 
This leads to an intuitively very attractive picture of interactions 
between guarks and glueballs, where quarks act much like certain 
enzymes act in the process of DNA replication by allowing one strand 
of the gluonic fluxtube to pass through another, thus changing its 
self-linking number and eventually leading to its decay 
into mesons. In particular, when quark loops are supressed
like in the limit of large quark masses or large-N, we expect the
knotted fluxtubes to be stable provided the space-time is 
four dimensional.
 
\vskip 0.5cm
In conclusion, we have studied the possibility that glueballs could 
correspond to closed fluxtubes. Very general universality
arguments then imply, that the stability of glueballs is possible
only if the string is twisted or knotted. Since the twist can be either
left-handed or right-handed, this implies that the
glueball spectrum must be degenerate, and in particular the
lowest mass glueball is a $0^{++}$ and $0^{-+}$ pair.
We have inspected the known meson spectrum up to 
energies around $2.3 \ GeV$, which is the upper limit that has been 
experimentally studied in $p \bar p$ annihilation processes. 
Our proposal that glueballs are closed gluonic fluxtubes 
then indicates, that the mysterious $\eta_L(1410)$ should
be interpreted as the $0^{-+}$ glueball, (essentially) 
degenerate in mass with the widely accepted
$0^{++}$ glueball $f_0(1500)$. We have explained how this interpretation
of glueballs as twisted and knotted strings arises in Yang-Mills
theory, where such configurations appear as knotted solitons in
an effective low-energy theory.
But our arguments are quite model-independent, 
suggesting that the interpretation of $\eta_L(1410)$ as the $0^{-+}$ 
glueball provides a test of various model independent
qualitative aspects of strong interactions. These include the general
properties of QCD string and confinement, the formation of a mass gap
in Yang-Mills theory, and the possibility of space-time to have 
more than four dimensions. Our estimate of the glueball mass 
spectrum suggests that these could all be experimentally
studied by the recently approved antiproton facility at GSI in Darmstadt.
\vskip 0.5cm
This article presents my talk at Quarks 2003, and I thank
the organizers for an excellent meeting. The article is largely based 
on original articles written together with Ludvig Faddeev
and Ulrich Wiedner, and I thank them both for wonderful collaborations.
The research has been partially supported by STINT IG2001-062
\vfill\eject

\end{document}